\documentclass[amssymb,twocolumn,aps,prstper,superscriptaddress,english,nofootinbib]{revtex4}
\usepackage{graphics,color,graphicx,amsmath}
\usepackage{subcaption}
\usepackage{natbib}
\usepackage{verbatim}
\usepackage{graphicx}
\usepackage[above,below]{placeins}
\usepackage{float}
\usepackage{tabularx}
\usepackage{times}
\usepackage{blindtext}
\usepackage{url}
\usepackage{rotating}
\usepackage{hyperref}

\renewenvironment{quote}
  {\list{}{\rightmargin=.1cm \leftmargin=.3cm}%
   \item\relax}
  {\endlist}

\begin{document}

\title{Instructing nontraditional physics labs: Toward responsiveness to student epistemic framing}

\author{Meagan Sundstrom}
\affiliation{Laboratory of Atomic and Solid State Physics, Cornell University, Ithaca, New York 14853, USA}

\author{Rebeckah K. Fussell}
\affiliation{Laboratory of Atomic and Solid State Physics, Cornell University, Ithaca, New York 14853, USA}

\author{Anna McLean Phillips}
\affiliation{School of Physics and Astronomy, Monash University, Clayton, Victoria 3800, Australia}

\author{Mark Akubo}
\affiliation{Laboratory of Atomic and Solid State Physics, Cornell University, Ithaca, New York 14853, USA}

\author{Scott E. Allen}
\affiliation{Department of Physics, Cornell University, 245 East Avenue, Ithaca, NY, 14853}

\author{David Hammer}
\affiliation{Department of Education, Tufts University, 12 Upper Campus Road, Medford, Massachusetts 02155, USA}
\affiliation{Department of Physics \& Astronomy, Tufts University, 574 Boston Avenue, Suite 304, Medford, Massachusetts 02155, USA}

\author{Rachel E. Scherr}
\affiliation{Physical Science Division, School of STEM, University of Washington Bothell, Bothell, WA 98011, USA}

\author{N. G. Holmes}
\affiliation{Laboratory of Atomic and Solid State Physics, Cornell University, Ithaca, New York 14853, USA}

\date{\today}

\begin{abstract}

Research on nontraditional laboratory (lab) activities in physics shows that students often expect to verify predetermined results, as takes place in traditional activities. This understanding of what is taking place, or \emph{epistemic framing}, may impact their behaviors in the lab, either productively or unproductively. In this paper, we present an analysis of student epistemic framing in a nontraditional lab to understand how instructional context, specifically instructor behaviors, may shape student framing. We present video data from a lab section taught by an experienced teaching assistant (TA), with 19 students working in seven groups. We argue that student framing in this lab is evidenced by whether or not students articulate experimental predictions and by the extent to which they take up opportunities to construct knowledge (epistemic agency). We show that the TA's attempts to shift student frames generally succeed with respect to experimental predictions but are less successful with respect to epistemic agency. In part, we suggest, the success of the TA's attempts reflects whether and how they are responsive to students' current framing. This work offers evidence that instructors can shift students' frames in nontraditional labs, while also illuminating the complexities of both student framing and the role of the instructor in shifting that framing in this context.  
\end{abstract}

\maketitle

\section{Introduction}

Physics laboratory (lab) instruction is shifting away from traditional, model-verifying activities that aim to reinforce content and toward open-ended investigations that engage students in practices of scientific experimentation~\cite{holmes2018introductory,smith2021best}. These practices include testing physical models with appropriate data collection methods, refining and iterating an experimental design, and deciding how to evaluate models using experimental data. Previous work demonstrates that students in nontraditional labs can engage in these practices ~\cite{holmes2015teaching,zwickl2015reasoning}, however many students' \textit{epistemic framing} -- their expectations of what knowledge they are supposed to produce in the lab and how they are supposed to produce it -- differ dramatically from the pedagogical design. Specifically, students often enter nontraditional labs with a \textit{confirmation} framing, expecting their experiment to verify a known theory or model that they learned in lecture~\cite{hu2017qualitative,hu2018examining,smith2018surprise,smith2020pretend,stein2018confirming,smith2020expectations}. Other work identifies a \textit{hoops} framing (as in ``jumping through hoops") in which students’ main goal is to finish the assignment and leave the lab as soon as possible~\cite{phillipsnotengaging}.

Research has shown that students in both the confirmation and hoops frames tend to engage in \textit{questionable research practices}~\cite{smith2020expectations,stein2018confirming,holmes2013doing,bogdan2013effects}, including with respect to how they handle unexpected experimental results~\cite{phillipsnotengaging}. Thus, one might argue that the instructional context of the lab, including the instructor behaviors during lab, ought to attempt to shift students out of these frames. Prior work has demonstrated that instructors can shift student frames in real time in problem solving contexts~\cite{modir2017students,irving2013transitions,thompson2016algorithmic,chari2019student}, yet instructor attempts to shift student frames in nontraditional labs are not always successful~\cite{phillipsnotengaging}. The primary goal of the current study, therefore, is to characterize ways in which instructors of nontraditional labs may shape student framing.

We performed a video analysis of seven student groups in one lab context: an early-semester, nontraditional lab activity in which the lab instructions were intentionally designed to shift student framing away from confirmation. The instructor of this particular session also explicitly confronts the confirmation frame by introducing the idea of \textit{falsification}, a concept of Popperian inquiry where an experimenter tests a hypothesis or claim that should be able to be refuted (i.e., falsified)~\cite{popper2013all,helfenbein2005falsifications,ladyman2012understanding,rowbottom2011kuhn}. The instructional context's attention to student framing, therefore, render this session ripe for analysis. In our study, we aimed to understand how students frame the lab activity and how the instructor shapes and cues that framing, with a particular attention to moments when student frames shift.


We characterize a set of student epistemic frames defined along two dimensions: whether or not students have a prediction for their experimental result and the extent to which students use their experiments to construct knowledge (epistemic agency). We observe that while the instructor shifts five out of the seven lab groups to take up a frame related to falsification, students do not take up instructor cues to frame the activity as one in which they have authority over knowledge construction. We argue that the instructor cues to shift student framing and behaviors are more often successful when they are \textit{responsive} to students' current framing. At the same time, however, even responsive cues might inadvertently lessen student agency.


\section{Theoretical Framework}

\begin{figure}[t]
\includegraphics[width=2.7in,trim={7.5cm 2.5cm 7.5cm 0cm}]{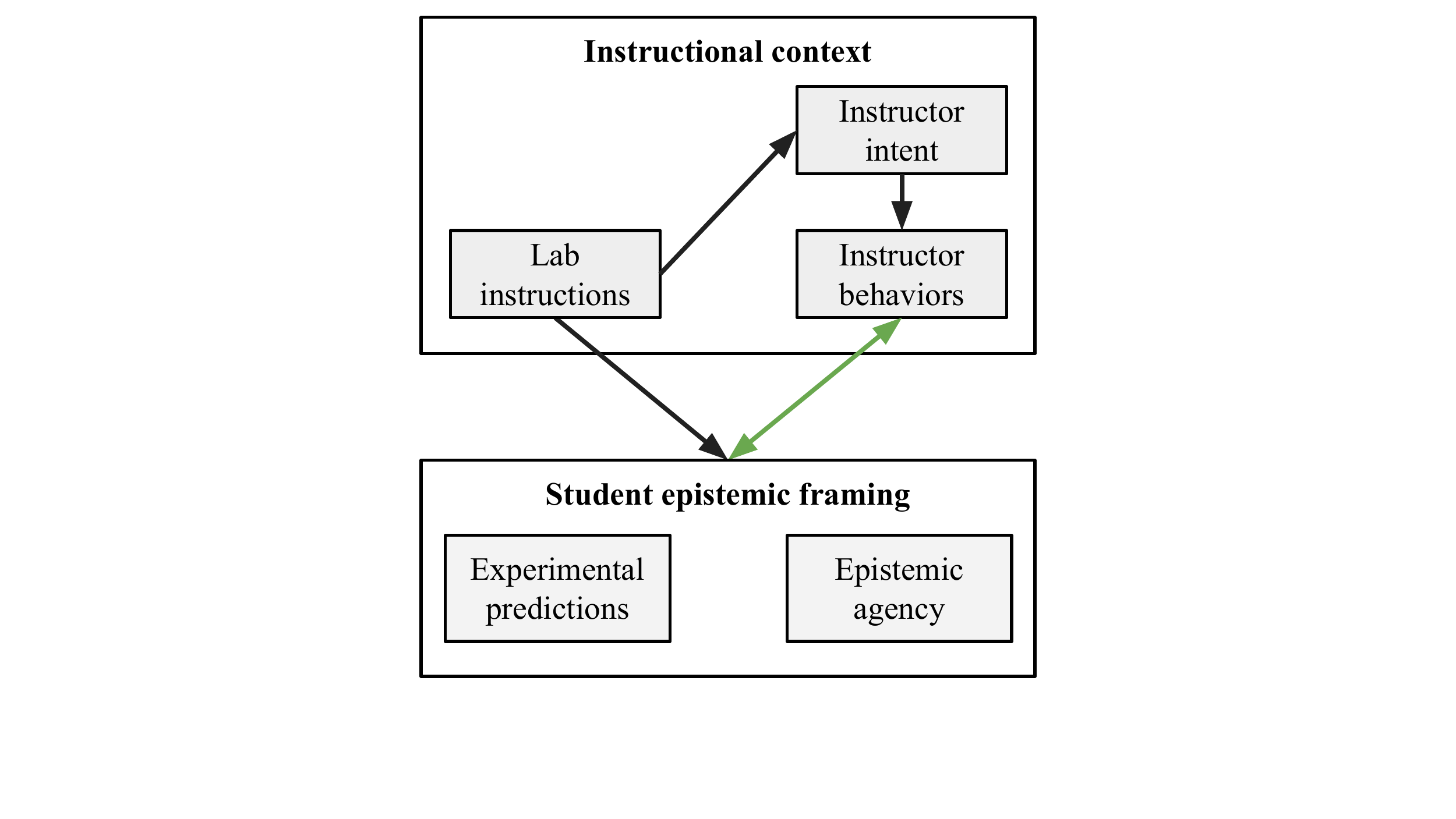}
\caption{Theoretical framework for this study. The instructional context encompasses components of the lab learning environment, including the written lab instructions, the instructor's intentions for teaching the lab, and the instructor's behaviors during the lab. The instructional context interacts with student epistemic framing, their expectations for what is taking place with regard to knowledge construction. This framing manifests in students' predictions for their experimental result and the extent to which they take up opportunities to construct knowledge (epistemic agency). Our main goal is to characterize the relationship between instructor behaviors and student framing (green arrow).
}
\label{theoreticalframework}
\end{figure}

We conducted our study using a theoretical framework that relates components of the instructional context to student epistemic framing (Fig.~\ref{theoreticalframework}). The primary goal of our analysis is to characterize the ways in which instructor behaviors may both cue and be responsive to student framing (green arrow in Fig.~\ref{theoreticalframework}). We further describe each relationship in the diagram below.

\subsection{Student epistemic framing}

The central component of our theoretical framework is student epistemic framing (bottom box of Fig.~\ref{theoreticalframework}). \textit{Framing} refers to how an individual or group interprets what is taking place or how they would answer the question, ``What is it that's going on here?"~\cite{goffman1974frame,tannen1993framing,TannenWallat}. \textit{Epistemic} framing, therefore, is an individual's or group's expectations related to what kinds of knowledge they are to construct and how they will construct it~\cite{shaffer2006epistemic, hutchison2010attending,redish2004theoretical, elby2010epistemological}. How students frame an activity impacts how they behave and engage with that activity. 


Researchers have investigated students' epistemic framing in a range of instructional science contexts, such as K-12 model-building activities~\cite{louca2004epistemological,rosenberg2006multiple,berland2012framing,shim2018framing,wendell2019epistemological,schellinger2022harmonious}, undergraduate chemistry labs~\cite{dekorver2015general,keen2022qualifying}, and undergraduate physics discussions~\cite{hutchison2010attending, scherr2009student,nguyen2016dynamics, modir2017students, modir2019framing} and labs~\cite{murphy2011answer,holmes2013doing,smith2018surprise,stein2018confirming,smith2020pretend,smith2020expectations,may2021students,phillipsnotengaging,bogdan2013effects}. Research on student framing in the latter context has demonstrated that students often frame their introductory physics labs as exercises in confirmation, such that they expect to verify a known theory or model through their experiments.  Confirmatory expectations have been identified across surveys of students' views of experimental physics~\cite{sneddon2009perceptions,leung2016students,hu2017qualitative,hu2018examining}, video analysis of students in labs~\cite{smith2018surprise, phillipsnotengaging,smith2020expectations}, and qualitative analysis of students' written lab notes~\cite{smith2020expectations,stein2018confirming,murphy2011answer,holmes2013doing}. Confirmation framing generally leads to problematic behaviors in the lab. For example, students holding confirmatory expectations tend to rely on the instructor or the lab manual as means of constructing knowledge because they believe the outcome of their experiment is predetermined by authority~\cite{smith2018surprise,smith2020expectations}. Moreover, students aiming to confirm a model may engage in questionable research practices -- ``decisions or behaviors that call into question the objectivity of experimental results"~\cite[p. 3]{smith2020expectations}. Such practices typically seek to align results with predictions from a model, such as through data manipulation, inflating or reducing uncertainty values, misinterpreting data to obtain particular results, and qualitatively judging results~\cite{smith2020expectations,stein2018confirming,holmes2013doing,bogdan2013effects}.~\footnote[1]{In some instances, confirmation framing can be productive~\cite{descamps2022perc}.}  

In the context of lab instruction, research suggests that student framing likely manifests in their articulation of \textit{experimental predictions} and in their enactment of \textit{epistemic agency}. Experimental predictions refer to whether or not students articulate an expected result for their experiment. Epistemic agency relates to the extent to which students expect to construct knowledge from their experiments~\cite{miller2018addressing,stroupe2014examining,zwickl2015reasoning}. For example, students in a confirmation frame will likely articulate an experimental prediction, such as that the experiment will lead to a result in line with a physical model they learned in lecture~\cite{smith2020expectations}. Such students may also exhibit low epistemic agency through behaviors such as failing to seek falsifying evidence~\cite{smith2020expectations} or ignoring data that disagree with their prediction~\cite{phillipsnotengaging}.

In our theoretical framework, students' experimental predictions and epistemic agency may or may not be related (no arrow between experimental predictions and epistemic agency in Fig.~\ref{theoreticalframework}), as students' enactment of epistemic agency does not necessarily depend on whether or not they expect a particular experimental result. For example, students holding a prediction may enact little epistemic agency by carrying out behaviors that will lead to confirming their prediction~\cite{smith2018surprise,smith2020expectations,phillipsnotengaging,zwickl2015reasoning}. Students taking up more epistemic agency, on the other hand, may believe that their prediction can be altered in the face of conflicting evidence, indicating a different framing~\cite{ProblematizingPERC,sundstrom2022perc,etkina2002role}. Furthermore, students with no prediction in mind may take up less epistemic agency if they do not engage in knowledge construction at all (e.g., the hoops frame~\cite{phillipsnotengaging}) or they may take up more epistemic agency by using their experiment to generate new knowledge~\cite{ProblematizingPERC}.




\subsection{Dynamic role of the instructional context}

Various components of the instructional context -- including the lab instructions, instructor intent, and instructor behaviors -- may directly or indirectly cue student framing. 

The lab instructor is a key component of the instructional context of the lab. An individual instructor constructs their intent for teaching the lab using a variety of resources, including the written lab instructions (arrow pointing from the lab instructions to instructor intent in Fig.~\ref{theoreticalframework}), as well as from other factors, such as their previous teaching and learning experiences and professional development. During instruction, an instructor may construct or amend their intentions based on what they notice about student participation. Indeed, research has demonstrated that individual instructors may implement the same instructional materials quite differently~\cite{Wan2020TA, McNeill2018, Spike2016, Goertzen2010Beliefs, Dancy2016}, bringing different expectations or frames to the science courses they teach~\cite{goertzen2010tutorial,wendell2019epistemological} and noticing different aspects of students' work and engagement~\cite{russ2013inferring}. 

The instructor enacts this intent through their behaviors during lab (arrow pointing from instructor intent to instructor behaviors in Fig.~\ref{theoreticalframework}), including their interactions with the students. These behaviors may directly cue student epistemic framing~\cite{modir2017students,irving2013transitions,thompson2016algorithmic,chari2019student} (arrow pointing from instructor behaviors to student framing in Fig.~\ref{theoreticalframework}); however, instructor cues to shift student frames may not always be successful~\cite{phillipsnotengaging}. This process also creates an indirect pathway where the lab instructions indirectly impact student epistemic framing through the instructor's intent and behaviors. 




The instructor may affect student framing through general features of how they run the lab (e.g., when introducing an activity) and in the particular ways they respond to what they notice in moments of interaction with students (arrow pointing from student framing to instructor behaviors in Fig.~\ref{theoreticalframework}). For example, if the instructor identifies that students are in a certain frame the instructor might tailor their interactions with the students to shift them to a different frame~\cite{chari2019student,robertson2016responsive}. We note that this identification does not require that the instructor understand ``framing" or ``frames" as a construct -- instructors may identify and respond to how students are understanding what is taking place without formal training that references student framing. 

\begin{figure*}[t]
\includegraphics[width=6in,trim={3cm 1.7cm 3cm 2cm}]{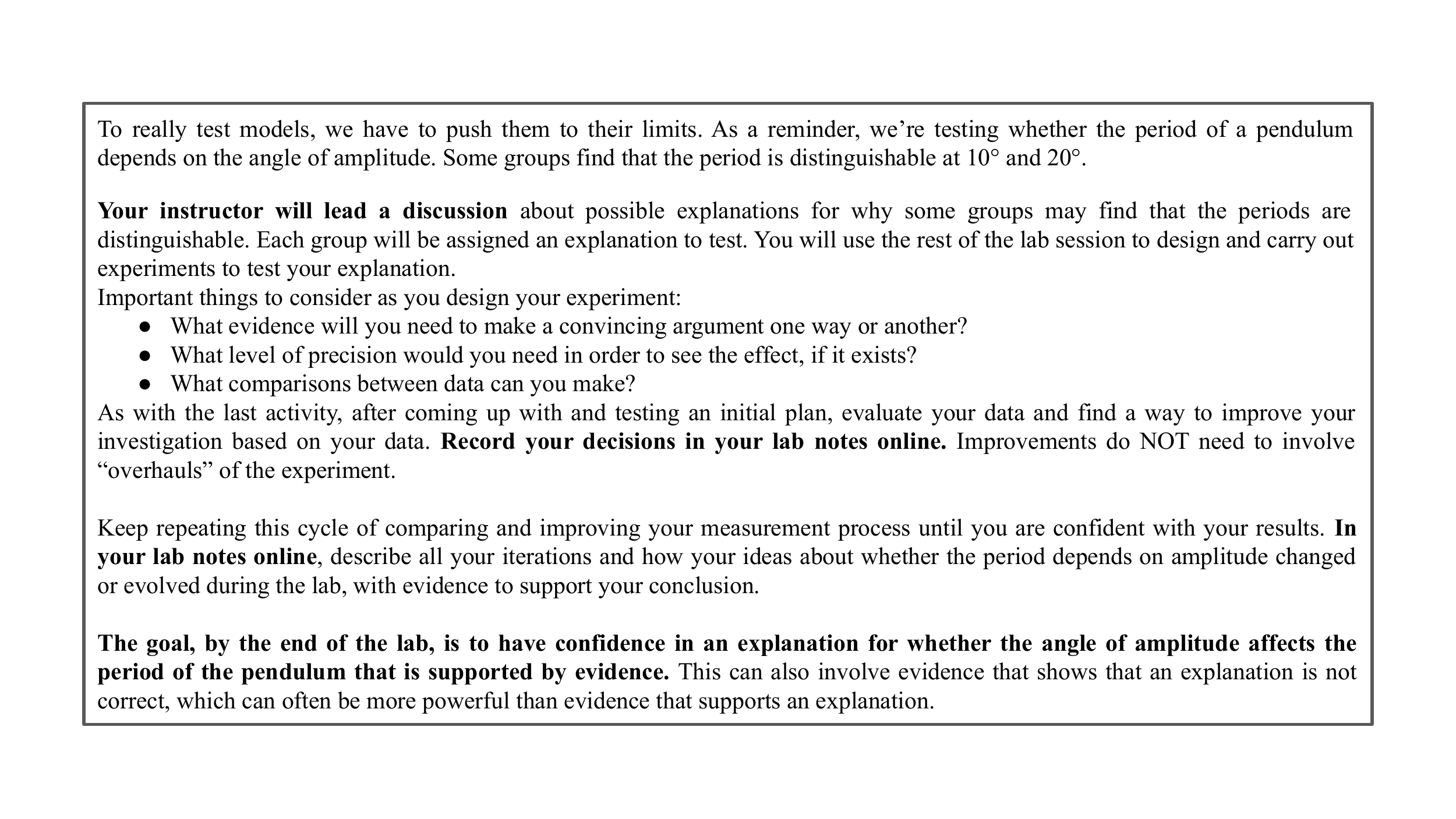}
\caption{The written lab instructions provided to students and the instructor for the mechanism testing activity.}
\label{labmanual}
\end{figure*}

Though we focus our analysis on the relationship between the lab instructor and student framing, the lab instructions (see Fig.~\ref{labmanual}) may also directly cue student epistemic framing (arrow pointing from lab instructions to student epistemic framing in Fig.~\ref{theoreticalframework}). For example, instructions that tell students exactly how to carry out their experiment and exactly what result they should expect to obtain may cue a confirmation framing in which students do not make their own experimental decisions~\cite{holmes2020developing}. 


\subsection{Summary}

We use this theoretical framework first as a grounding for identifying and characterizing a broader set of student epistemic frames in nontraditional physics labs. Second, we use it to understand the complex relationship between the instructor's behaviors and student epistemic framing. Below, we characterize the data analyzed in this study for each of the boxes in Fig.~\ref{theoreticalframework}. These data include analysis of the lab instructions provided to the students and instructor, an interview with the instructor to illuminate their intentions, and video data of student and instructor behaviors during a nontraditional physics lab.

\section{Instructional context}

The data in this study come from one lab section of an introductory, calculus-based mechanics course at Cornell University. This section contained 19 students, 11 men and eight women. Most students identified as White and/or Asian or Asian American, and the majority were first- or second-year students intending to major in engineering. The teaching assistant (TA) for this section was a graduate student familiar with physics education research (we omit the TA's demographic information to preserve anonymity). 

Students in this physics course attend one 2 hr lab session each week. 
The labs aim to engage students in practices of scientific experimentation~\cite{holmes2018introductory,smith2020direct,holmes2015teaching}. Lab activities provide context (e.g., a physical phenomenon and models to describe it) but leave many decisions up to the students (e.g., how to set up their apparatus, how many data points to collect, and how to evaluate the validity of the provided models). Students work in lab groups of two to four and submit lab notes as a group, which are then graded by the TA. 

There are four main lab units that span two weeks each. In the first unit, students test whether the period of a pendulum depends on the angle from which it is released. During the second unit, students analyze the acceleration of objects in free fall to determine the forces acting on the object. In the third unit, students design experiments using stretchy objects to test the assumptions and limitations of Hooke's Law. The final unit is a project lab in which students extend their investigation from one of the previous units. The lab instructions for all units are available on PhysPort.org~\cite{PhysPortLabs}.

We focus our analysis on the final activity of the first unit -- the pendulum lab. We selected an early-semester lab because students likely hold a confirmation frame at this point~\cite{smith2020expectations} and because the instructional materials explicitly attend to student framing, described next. 

In the first week of the pendulum unit, students measure and compare the period of a pendulum when released from 10$^{\circ}$ and 20$^{\circ}$. Most of the time is spent identifying and reducing sources of uncertainty in the measurement process. The instructor introduces a statistical measure called \textit{t-prime} ($t'$)~\cite{holmes2015quantitative}: $t'=\frac{A-B}{\sqrt{\sigma_A^2+\sigma_B^2}}$, where $A$ and $B$ are measurements, and $\sigma_A$ and $\sigma_B$ are their uncertainties.  Students use \textit{t'} to determine the distinguishability of their two data sets (one per angle). This measure is similar to a Student's t-test, where the following outcomes and interpretations are possible:
\begin{itemize}
\itemsep0em 
    \item $t'<1$: the data sets are indistinguishable, the period of a pendulum when released from 10$^{\circ}$ and 20$^{\circ}$ is likely the same;
    \item $1<t'<3$: inconclusive, we do not have enough statistical power to tell whether the period of a pendulum when released from 10$^{\circ}$ and 20$^{\circ}$ is the same or different;
    \item $t'>3$: the data sets are distinguishable, the period of a pendulum when released from 10$^{\circ}$ and 20$^{\circ}$ is likely different.
\end{itemize} 
Our study focuses on the second week of the lab activity. Students spend the first half of the session further reducing sources of uncertainty from the previous week and comparing the periods from 10$^{\circ}$ and 20$^{\circ}$. In many cases, students' precise measurements lead them to a discrepancy: the period measurements from 10$^{\circ}$ and 20$^{\circ}$ become distinguishable. This result contradicts the model of a pendulum typically given in introductory texts, $T = 2\pi \sqrt{l/g}$, which suggests no dependence on angle of release. Typically, at least one group in a given lab section finds a significant difference in the period from the two angles ($t'>3$). 

In the section we study, one group finds this discrepancy, and the TA draws on their result to motivate the follow-up ``mechanism testing'' activity (about 1 h long), which we analyze (see Fig.~\ref{labmanual}). In this activity, students brainstorm and design experiments to test mechanisms (e.g., related to the measurement process or the physical model itself) that could explain why that group may have observed the discrepancy. For example, a group might conduct an experiment to determine if the discrepancy is due to air resistance acting against the motion of the pendulum.

\subsection{Lab instructions}
\label{sec:instructions}

The planned flow of the lab as a whole assumes both that most students expect the lab to confirm there is no dependence on amplitude and that the students' data show a dependence on amplitude. The follow-up mechanism testing activity (Fig.~\ref{labmanual}) guides students to interrogate this result that conflicts with their expectations. Thus, the instructions assume that students approach this activity with a confirmation framing, given prior research  ~\cite{sneddon2009perceptions,leung2016students,hu2017qualitative,hu2018examining,smith2018surprise, phillipsnotengaging,smith2020expectations,stein2018confirming}, and the instructions aim to confront this framing. For example, the instructions speak directly of how ``Some groups find that the period is distinguishable at 10$^\circ$ and 20$^\circ$," and also direct students to address this mismatch between findings and expectations. This, we note, might present an epistemological tension: if students do not see the discrepancy in their own data, they may not be motivated to interrogate the discrepancy found by other groups in the subsequent activity. 

The mechanism testing activity instructions prompt students to design an experiment to test whether a given mechanism is a plausible explanation for the observed discrepancy: ``The goal, by the end of the lab, is to have confidence in an explanation for whether the angle of amplitude affects the period of the pendulum that is supported by evidence." These instructions prepare students for the possibility of finding any experimental result, so long as it is supported by their data. The instructions also, however, emphasize finding evidence that disproves, rather than proves, that their assigned mechanism is a plausible explanation for the angle dependence: ``This can also involve evidence that shows that an explanation is not correct, which can often be more powerful than evidence that supports an explanation." This sentence may reflect a tension in the instructions between framing the experiment as open-ended and as aiming for falsifying evidence (i.e., evidence that falsifies explanations for why the period of the pendulum may depend on the angle of release). 

This tension is related to epistemic agency, as the instructions simultaneously constrain and support such agency. While most curricula both constrain and support student agency, the inconsistent instructions here might prompt different student behaviors. On one hand, the instructions limit agency by promoting falsifying evidence over other evidence and eliminating students' freedom to choose what they think is most appropriate mechanism to test: ``Each group will be assigned an explanation to test." On the other hand, students are prompted to make their own experimental decisions and iterate on their design: ``After coming up with and testing an initial plan, evaluate your data and find a way to improve your investigation based on your data."  Similarly, the instructions maintain ambiguity about the experimental result [e.g., ``have confidence in an explanation for \emph{whether} the angle of amplitude affects the period..."(emphasis added)]. 

In sum, the instructions reflect pedagogical goals of disrupting confirmation framing and supporting students taking up epistemic agency. At the same time, different parts of the instructions may cue different student frames with respect to how they perceive themselves in using their experiment to construct knowledge. The instructions may also inform the instructor's intent, which we discuss below.

\subsection{Instructor intent}
\label{sec:interview}

As mentioned, the instructor of this particular lab section was familiar with physics education research and the literature related to student framing in instructional labs. They also had multiple years of experience teaching university-level physics. The instructor, therefore, likely used this experience and body of knowledge in addition to the lab instructions to construct their intentions for teaching the lab. 

To better understand the instructor's intent, the first author conducted a semi-structured, virtual interview with the TA of the analyzed lab session. The interview took place several years after the analyzed session occurred. While the TA may not have remembered specific details of this particular session, they likely recalled teaching the labs more broadly. This instructor also did not teach a fully in-person version of these labs again after the semester we analyzed, so their most recent, relevant experience was during the semester that the data for this study was collected.

The goal of the interview was to understand whether and how the TA intended to cue student framing in the mechanism testing activity. The interview protocol (provided in the Appendix) asked the TA to describe their previous teaching experience, their understanding of the instructional goals of the mechanism testing activity, and how they attended to student framing when teaching this activity.

In the interview, the TA expressed that they intentionally attended to and cued student epistemic framing when teaching this lab course more generally: 
\begin{quote}
\emph{[Students] are like ‘Well, if I’m not supposed to be coming up with the correct answer and I’m not supposed to be showing you that my idea was a good idea, what am I supposed to be doing?’, and then you, you kind of help them discover what they are supposed to be doing, and that’s fairly difficult, right, [to present] this really consistent picture of ‘What am I supposed to be doing? What am I supposed to be producing?’}
\end{quote}
The TA described that students typically hold a confirmation framing during this activity: they have an incoming expectation to find and provide a ``correct answer.'' In response, when instructing this lab, the TA tried to directly confront this confirmation framing by reshaping students' ideas of what they are ``supposed to be doing" and ``supposed to be producing," that is, reshaping their epistemic framing.

The TA also described that to do this reshaping, they introduced the concept of \textit{falsification} -- using an experiment to test a claim that is able to be disproved or falsified:
\begin{quote}
\emph{I know at some point on the board, I talked about it: with this relation to the model, falsify the model, find where the model fails. But at some point I wrapped that into like ‘Figure out which part of your experiment isn't working and revise that.'}
\end{quote}
The TA introduced falsification as a broader epistemology for doing science, both as a way of model testing -- trying to find if and when ``the model fails" rather than showing limited evidence that the model is correct -- and as a means of experimental design -- identifying what experimental choices need improvement. These intentions likely stemmed directly from the lab instructions, which aimed to confront students' desire find a correct experimental result and focus students on reducing uncertainty in their experimental design.

We also asked the TA in the interview how they expected students to operationalize the idea of falsification in their lab experiments. The TA described that they intended for students to design an experiment capable of finding the limitations of a model, but not to \textit{necessarily} falsify the model (ellipses indicate omitted speech):
\begin{quote}
\emph{``The only way that some experiment shows ‘Oh, yeah we prove this theory,’ the only way that that experiment is meaningful is if it was capable of proving it wrong...So do yourself a favor and design an experiment that’s capable of doing that, but it's not like you have some obligation, or I have some expectation, that you interpret the data -- that like you squint and turn your head to like make the data say something it doesn't say."}
\end{quote}
The TA, therefore, viewed experimental results as open-ended: they expected students to design an experiment capable of falsifying a model, but wanted students to appropriately interpret their data rather than find any particular result. 

To summarize, the TA used the lab instructions, their previous knowledge, and their previous experiences to construct their intent of confronting students' incoming expectations of confirmation. They did so by introducing a new falsification framing that was meant to encourage students to try to find (but not necessarily find) whether and to what extent a given model or claim under investigation is limited, while also iteratively reflecting on and improving their experimental designs.


\section{Methods}

\subsection{Data Collection}

We collected video and audio data of this lab section during the mechanism testing activity because we can infer individuals' framing from their speech, gestures, and behaviors~\cite{scherr2009student,scherr2009video,berland2012framing}. We placed two wall-mounted video cameras in opposite corners of the classroom to capture the physical behaviors of all seven groups in the section. We also placed an audio recorder on each table to capture clear audio recordings of individual lab groups. The TA wore a separate audio recorder such that we had a complete audio track of their speech. To sync the recordings, we paired the TA's and each group's audio track with the video recording that best captured each group.

\subsection{Analysis}

Similar to our prior work analyzing student discourse in labs~\cite{phillipsnotengaging}, one author watched the video of each of the seven lab groups in the session in its entirety without pausing or writing anything down. On the second watch, the author made notes summarizing the group's actions at five minute intervals. In these summaries, they also noted and partially transcribed moments when students were discussing their expectations for the experiment (i.e., when there was evidence of epistemic framing). All of the videos were fully transcribed because many of these moments needed context from other parts of the activity to fully understand and analyze. 

\begin{figure*}[t]
\includegraphics[width=5.8in,trim={4cm 1cm 4cm 0.5cm}]{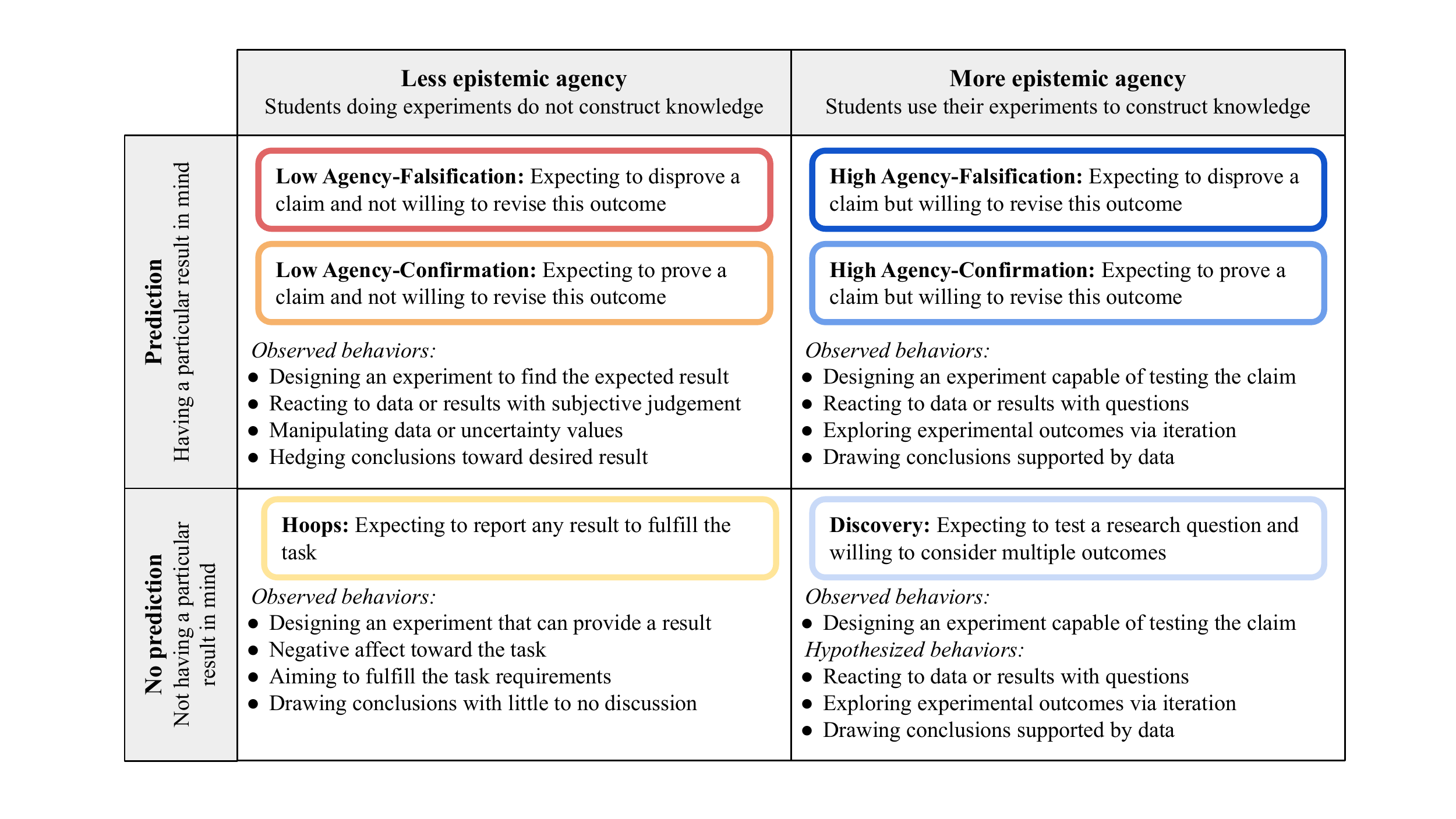}
\caption{Definitions of and observed behaviors corresponding to students' epistemic frames identified from the video analysis. Frames are shown along two dimensions -- predictions and epistemic agency -- in line with our theoretical framework (see Fig.~\ref{theoreticalframework}). Students were rarely in the Discovery frame (see Fig.~\ref{timeline}), therefore we did not observe many behaviors corresponding to this frame. Instead, we propose \textit{hypothesized} behaviors based on our observations of the High Agency-Confirmation and High Agency-Falsification frames.}
\label{frames}
\end{figure*}

The first and second authors then selected short clips of the videos they identified as salient moments related to students' epistemic framing and brought them to larger group discussions among the research team. In these discussions, we talked about both the students' and the instructor's epistemic framing and other aspects of the students' discourse that stood out to us. From this set of episodes, the research team inductively and iteratively developed a list of clusters of student speech and behaviors. We associated these clusters with particular student epistemic frames (shown in Fig.~\ref{frames}), which we classified based on students' experimental predictions and epistemic agency per our theoretical framework (see Fig.~\ref{theoreticalframework}). 

With these frames identified, the first and second authors individually coded the videos for each lab group, including both the students' framing and the TA's framing during their interactions with the group, from the point at which the group began discussing the activity until they left the room. The videos were coded using the list of epistemic frames at the level of communicative events: portions of dialogue where the topic of conversation and general activity remained constant~\cite{hennessy2016developing, derry2010conducting}. Moments when student activity did not seem related to a particular frame or when the two authors were together unable to assign a frame were marked with the code ``unclear." These moments generally consisted of times when students began or switched activities and there was insufficient evidence of any particular framing. Moments when the instructor spoke to students about available equipment, reflected on the activity at the end of the session as students were leaving, and participated in off-topic discussion were marked as ``not related to framing." Once their initial coding was conducted, the two authors met, discussed all video, and agreed on consensus codes. 

As frames are communicated both verbally and non-verbally among groups~\cite{tannen1993framing}, most of the time we coded a single frame for an entire group of students. In one case, we found evidence that two members of a group consistently had different frames, consistent with literature describing that epistemic frames may be evident in individual actions and behaviors as well as those of groups~\cite{scherr2009student}. This case was coded as two different frames.

We also identified and conducted more detailed interaction analysis~\cite{derry2010conducting, jordan1995interaction} of the moments leading up to, during, and following TA-student interactions.
Below, we present the results of our coding, which show broad patterns in student and instructor frames, and interaction analysis.

\section{Results}
\label{sec:results}

\begin{figure*}[t]
\includegraphics[width=7.2in,trim={0 3cm 0 0}]{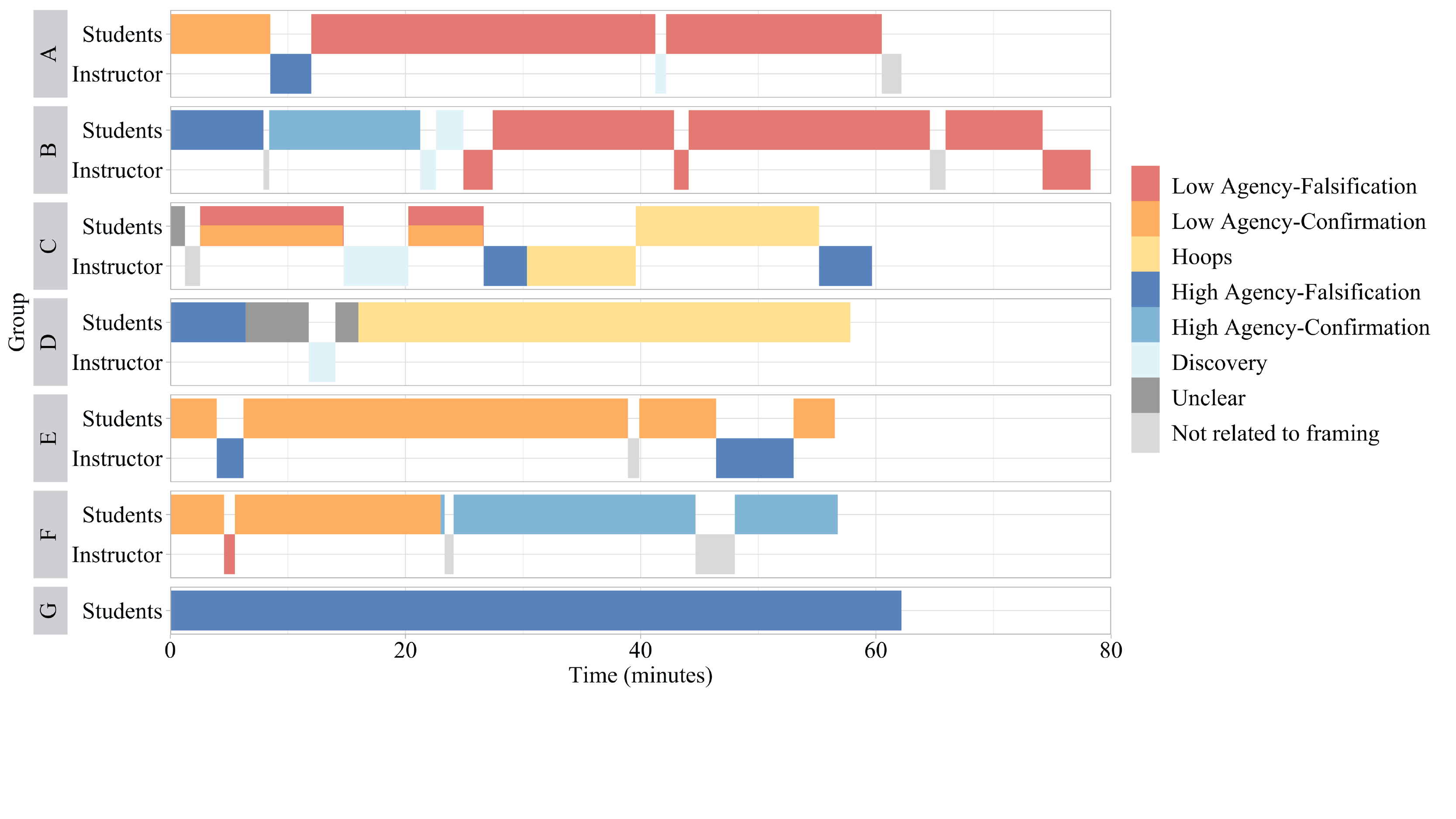}
\caption{Timeline figure depicting student and instructor framing (defined in Fig.~\ref{frames}) over time during the mechanism testing activity. Two simultaneous colors (i.e., in Group C) indicate that individual students in the same group were in different frames. The start time for every group (0 min mark) corresponds to when they start discussing the activity and the end time corresponds to when the group finishes the activity and leaves the room (hence the different lengths of time in each group and the apparent overlap of instructor interactions).}
\label{timeline}
\end{figure*}

As mentioned in the instructor interview (see Section~\ref{sec:interview}), the TA makes a pitch for falsification to the whole class. At the beginning of the mechanism testing activity, the instructor writes a list of mechanisms that may explain why the period depends on angle of release on the board and describes the assignment:
\begin{quote}
    \textit{``Okay, so now what I’m going to do is, each of you are going to investigate a different one of these. So what you want to think is, which of these are quantifiable and testable. Now again, another word for testable is falsifiable. Right, so you weren’t trying to prove, and you -- we’re not trying to prove that this equation is true. You’re trying to find where it fails. The only way to investigate something is to try to prove it wrong. So, as you design your experiment, you should not try to prove why the one you picked is the right one. You should be attempting to falsify it.”}
\end{quote}
This pitch is an attempt to directly confront students' confirmation framing: the TA provides a contrast between ``proving” a claim under investigation is correct and ``finding where it fails.'' Here, the claim being tested is that a given mechanism explains why one of the groups observed that the period of the pendulum depends on amplitude. The TA also explicitly cues the students to try to prove this claim wrong (e.g., ``You should not try to prove why the one you picked is the right one. You should be attempting to falsify it."). 

In the following sections, we describe students' epistemic framing after this pitch for falsification and the instructor's behaviors when interacting with individual lab groups during the activity.

\subsection{Student epistemic framing}

We identify six different epistemic frames (Fig.~\ref{frames}) and observe that each lab group takes up and shifts between these epistemic frames differently throughout the activity (Fig.~\ref{timeline}). Here we describe these frames and broad trends in student framing. 

Based on our theoretical framework (see Fig.~\ref{theoreticalframework}), we define students' epistemic frames along two dimensions (Fig.~\ref{frames}): whether or not students have a prediction for their experimental result and the extent to which students enact epistemic agency to construct knowledge from their experiment. We also distinguish the predictions themselves, whether confirmation or falsification.

Students holding a prediction for the experiment (i.e., verbalizing an expectation to either falsify or confirm their mechanism) but exhibiting little to no epistemic agency take up the \textbf{Low Agency-Falsification} frame or the \textbf{Low Agency-Confirmation} frame. Students in one of these two frames do not attempt to construct knowledge, rather they design an experiment that will yield their expected result and may manipulate their data or uncertainty values to push their result toward their prediction. One group (Group E) in the Low-Agency Confirmation frame, for example, verbalizes the purpose of their experiment as, ``\textit{The whole point is we’re supposed to fudge it to say yes.}” Students in one of these two frames also hedge the phrasing of their conclusions to more closely align with their prediction. One group (Group B) in the Low Agency-Falsification frame, for instance, discusses what to conclude, “\textit{Okay yeah let’s just say ‘may not’ then cause we’re trying to falsify it right?}” 

Students having a prediction for the experiment and enacting more epistemic agency, in contrast, take up the \textbf{High Agency-Falsification} frame or the \textbf{High Agency-Confirmation frame}. Students in one of these two frames use their experiment to construct knowledge: they design an experiment capable of testing the claim that their mechanism explains the observed angle dependence of the pendulum, but note the possibility of different outcomes and react to their data or results with questions. One group (Group B) in the High Agency-Confirmation frame, for example, expects to confirm the claim under investigation but remains open to other outcomes that will be determined by their statistical analysis. A group member notes, “\textit{So, after we take the trials, how are we proving that it affects the dependency?...Okay, so if $t'$ is small, then it doesn't affect it, but if it’s large it does.}” Another common behavior of students in these frames is drawing appropriate conclusions supported by the data rather than hedging results toward their prediction. One student (Group G) in the High Agency-Falsification frame, for example, interprets their data in light of their expectation of falsifying the claim: “\textit{I mean like for 20 degrees it kind of falsifies that claim, but the 10 degrees one makes it a little...}”

Other groups do not verbalize a predicted outcome for their experiment (bottom row of Fig.~\ref{frames}). Students with no prediction and enacting little epistemic agency take up the \textbf{Hoops} frame, similar to that identified in previous literature~\cite{phillipsnotengaging}. These students express a desire to complete their lab experiment quickly and often display negative affect toward the activity. Students in the Hoops frame also design an experiment that can provide any result (i.e., they are not constrained by a prediction). One group (Group C), for example, does not spend much time designing their experiment and instead views the experiment as an assignment they need to get done: \textit{“Let’s just get some numbers down.”} 

Students with no prediction and enacting more epistemic agency, on the other hand, take up the \textbf{Discovery} frame. Students in this frame have no expectation for the experimental result and design an appropriate experiment to test the claim,  remaining open to multiple possible results. One group (Group B) in the Discovery frame, for instance, outlines their goal as answering an open research question: \textit{“So we are trying to see if tension makes angle matter."}

Summing across all groups' framing trajectories (Fig.~\ref{timeline}), students spend about a third of the activity in the Low Agency-Falsification frame (red bars of student frames spread across Groups A, B, and C). On aggregate, students also spend about a fifth of the time in the High Agency-Falsification frame (dark blue bars of student frames spread across Groups B, D, and G). The prevalence of these two frames signify that the instructor's initial pitch for falsification did shape student framing, particularly with regard to experimental predictions: students were in a frame where they expected to falsify the claim under investigation for roughly half of the lab activity and five out of seven groups exhibited one of these two frames at some point during the activity.

In total, students also spend about half the activity in either the Low Agency-Falsification or Low Agency-Confirmation frames (red and orange bars of student frames spread across five of the seven groups, all but Groups D and G), explicitly seeking to falsify or confirm the claim under investigation. This is unsurprising given the literature on the dominance of the confirmation frame especially early on in the semester~\cite{smith2020expectations,smith2018surprise,smith2020pretend,phillipsnotengaging,stein2018confirming}, however this is counter to the goals of the lab and the instructor’s cue toward a more open-ended result in their falsification pitch. Similarly, two groups spend more than a third of their time in the Hoops frame (yellow bars of student frames in Groups C and D), expecting to conclude anything the data says for the sake of completing the assignment. While this is also unsurprising given the research literature~\cite{phillipsnotengaging}, it is also counter to the goals of the lab.

Students spend a lower proportion, about a third, of the total time across all groups in one of the three frames associated with more epistemic agency (three shades of blue bars of student frames spread across Groups B, D, F, and G). These frames are only dominant in two of the seven lab groups (Groups F and G).

\subsection{Instructor behaviors}


Here we summarize broad trends in the instructor's framing when interacting with individual student groups during the activity, shown in the Instructor rows of Fig.~\ref{timeline}. In the following section, we present the results of our interaction analysis.

We observe that the TA spends more time interacting with groups about how they are framing the activity than on other aspects of the activity (colored bars versus gray bars of instructor frame in Fig.~\ref{timeline}). The TA also predominantly (about 40\% of their total time spent interacting with students) attempts to shift students into the High Agency-Falsification frame (dark blue bars of instructor frame in Fig.~\ref{timeline}), aiming for students to try to falsify their assigned mechanism but remaining open to other outcomes supported by their data. This aligns with the TA's intent described in the instructor interview (see Section~\ref{sec:interview}). Similarly, the TA attempts to cue the Discovery frame for about a fifth of the time that they spend interacting with individual groups (light blue bars of instructor frame in Fig.~\ref{timeline}). 

The TA also, however, tries to cue the Low Agency-Falsification frame in a few interactions (red bars of instructor frame in Fig.~\ref{timeline}, summing to about a fifth of the time), suggesting to students that falsification is the experimental outcome they should find. Combined with the time spent cuing the High Agency-Falsification frame mentioned above, the TA spends roughly 60\% of the interaction time trying to shift students into a falsification-oriented frame (sum of dark blue and red bars in instructor frame in Fig.~\ref{timeline}). The TA never prompts the Low Agency-Confirmation or High Agency-Confirmation frames (no orange or medium blue bars in instructor frame in Fig.~\ref{timeline}), in line with their intent to confront students' expectations of confirming the claim under investigation (that a given mechanism explains the observed angle dependence).

Surprisingly, the TA cues the Hoops frame with Group C (yellow bar of instructor frame in Fig.~\ref{timeline}) despite this framing contradicting the goals of the lab. In this case, the mechanism that the group is testing is the breakdown of the small angle approximation. That is, their assignment is to test whether or not the observed difference in the pendulum period when released from 10$^{\circ}$ and 20$^{\circ}$ is attributable to the small angle approximation (which is used in formulating the canonical expression for the period of a pendulum) not holding at these angles. One group member verbalizes a desire to confirm that the mechanism is correct, while a second group member wants to disprove, or falsify, that the mechanism is correct (simultaneous red and orange bars of student frame in Fig.~\ref{timeline}). While negotiating this framing misalignment, the group struggles to design an experiment until late in the session (they do not collect data until more than 40 minutes into the activity) and visibly expresses frustration with the activity. To ensure the group has an experiment to report on before class ends, the TA tells them, “\textit{You don’t actually have to succeed. You just have to do an experiment that investigates the predictions that come from the hypothesis you’re investigating...In the setting we’re in now, that’s good enough.}”

\subsubsection{Student frame shifts cued by the instructor}

With the exception of Group G, all lab groups take up more than one epistemic frame during the course of the activity and therefore shift between different frames. Here we describe two examples of student frame shifts that are shaped by interactions with the instructor.


Despite starting the activity in different frames and receiving cues for different frames from the instructor, Groups A and B both spend the majority of their time in the Low Agency-Falsification frame (red bars for student frame in Fig.~\ref{timeline}). As we will show, the TA responds to both groups as if they are in a confirmation frame and tries to confront this frame by cuing either the High Agency-Falsification frame (Group A) or the Low Agency-Falsification frame (Group B). These TA moves are in line with their intentions clarified in the instructor interview, however the result is neither group framing the activity as one where they construct knowledge: both groups shift to the Low Agency-Falsification frame.

\textbf{Group A} -- Alex, Mike, and Faru -- is initially in the Low Agency-Confirmation frame (orange bar for student frame in Fig.~\ref{timeline}), aiming for their experiment to demonstrate that their mechanism, the three-dimensional (3D) motion of the pendulum swing (as oppposed to the pendulum swinging in a plane), is the correct explanation for the observed angle dependence. Alex proposes an experimental design:
\begin{quote}
    \textbf{Alex}: \textit{I think the idea is if with the 3D motion 10$^{\circ}$ and 20$^{\circ}$ are distinguishable, without the 3D motion they are indistinguishable, then we would have, is that the way to show that that is the source?}\\
    \textbf{Faru}: \textit{Yeah.}
\end{quote}
Alex suggests for the group to add more 3D motion to the pendulum's trajectory and again compare the periods from 10$^{\circ}$ and 20$^{\circ}$. If they find that the two periods differ in these conditions, Alex offers, they can ``show that that is the source" of the angle dependence. In other words, the group intentionally sets up an experiment to find their expected result.

When the group presents this idea to the TA, the TA confronts their intentions of confirmation by attempting to cue the High Agency-Falsification frame (dark blue bar for instructor frame in Fig.~\ref{timeline}):
\begin{quote}
\textbf{TA}: \textit{It’s harder to prove that it is the [3D motion], so you could also prove that it's not.}\\
\textbf{Alex}: \textit{That it's, yeah.}\\
\textbf{Mike}: [Inaudible]\\
\textbf{Alex}: \textit{I guess.}\\
\textbf{TA}: \textit{Much easier to prove.}\\
\textbf{Alex}: \textit{Much easier to prove that it's not the source of error.}\\
\textbf{TA}: \textit{Yeah.}\\
\textbf{Alex}: \textit{So that would basically mean, they're distinguishable in both cases 10$^{\circ}$ and 20$^{\circ}$ are both distinguishable, the angle-related error is present in both cases.}\\
\textbf{TA}: \textit{I'm gonna give you a hint. You don't even necessarily have to check at different angles. Right, so the hypoth-- the 3D motion hypothesis goes that in a perfect world there is no angular dependence, but by increasing the angle you also have a tendency to increase this kind of [non-linear] motion and that the angular dependence you think you're observing is indeed just a circular motion dependence. So try to measure the effect of circular motion on the period and see if it fails to explain the extent of the angular dependence we saw previously.}\\
\textbf{Alex}: \textit{Right, and you can just, I guess, test that with 10$^{\circ}$ and see if the period changes with 10$^{\circ}$ and that would be evidence that}\\
\textbf{TA}: \textit{Yeah if you did that only at 10$^{\circ}$, you did a couple conditions all at the same angle, but something having to do with 3D motion. Right, so there are a couple ways you could do it, you could eliminate it and see what happens, right?}\\
\textbf{Alex}: \textit{Awesome, okay.} [TA  leaves table] \textit{I did have this one idea, so} [inaudible] \textit{the 3D motion} [TA returns]\\
\textbf{TA}: \textit{But again you're not gonna get graded on whether or not it turns out you're right. So you're gonna be graded on whether your justifications for your conclusions are good and whether you used proper statistics as you went, and so forth.}
\end{quote}
In this interaction, the TA tries to cue the High-Agency Falsification frame in a few ways. First, the TA suggests that falsifying a claim is easier than proving it correct (e.g., ``It's harder to prove that it is the [3D motion], so you could also prove that it's not."). The TA also offers direct suggestions for the students to set up an experiment capable of falsifying the claim (e.g., ``You don't even necessarily have to check at different angles.") and states that there is no particular result they need to find (e.g., ``See what happens."). Finally, the TA notes that their grade does not depend on the experimental result itself (e.g., ``You're not gonna get graded on whether or not it turns out you're right."). Together, these moves comprise an attempt to cue the High Agency-Falsification frame in which the students might expect to find falsifying evidence, but ultimately draw appropriate conclusions supported by their data. 

After this interaction with the TA, the group shifts to the Low Agency-Falsification frame (red bar for student frame in Fig.~\ref{timeline}). The students take the TA’s suggestions to mean that they should do an experiment where the result is that their mechanism does not explain the angle dependence (i.e., falsifying the claim under investigation). They perform the experiment suggested by the TA (i.e., compare the period when released from 10$^{\circ}$ with and without adding extra 3D motion) and discuss their incoming data: 
\begin{quote}
\textbf{Alex}: \textit{We’re trying to falsify this. Alright, ready?} [releasing the pendulum]\\
\textbf{Faru}: \textit{Yeah. 1.27}\\
\textbf{Alex}: \textit{Aw, that’s good, that’s pretty good.}\\
\textbf{Mike}: \textit{We have to falsify this.}\\
\textbf{Alex}: \textit{Hm?}\\
\textbf{Mike}: \textit{How is this gonna falsify it?}\\
\textbf{Alex}: \textit{Basically saying like this is not the source of error, this is not the source of difference, this is not the extra} [inaudible]\\
\textbf{Mike}: \textit{Cause right now it looks like it’s not gonna falsify it.}
\end{quote}
As they collect the period measurements, the group reacts subjectively to the data (e.g., ``That's good.") and Mike expresses concern that they will not obtain their desired result of falsification (e.g., ``Cause right now it looks like it’s not gonna falsify it."). These comments exemplify the group's aim to find the result they think the TA wants them to find, and they remain closed off to finding a different outcome.


To summarize Group A's trajectory, the TA attempts to cue the High Agency-Falsification frame upon identifying that the students are initially in a confirmation frame. This interaction shifts the group from expecting a confirmatory result to expecting a falsifying result, but does not shift the group from low to high epistemic agency (the groups shifts to the Low Agency-Falsification frame).

\textbf{Group B} -- Sarah, Alina, and Carlos -- tests whether the tension in the pendulum string explains the observed angle dependence of the pendulum period (e.g., maybe there is more tension at one angle than the other). About ten minutes into the activity, the group is in the High Agency-Confirmation frame (medium blue bar for student frame in Fig.~\ref{timeline}). They decide to compare the period of the pendulum released from 10$^{\circ}$ and 20$^{\circ}$ with a heavier bob than the previous experiment and discuss their expectations for the result:
\begin{quote}
\textbf{Sarah}: \textit{So, um, basically what I’m thinking is we could do 5 or 10, what do you guys think? Like 5 or 10 trials with the weight and then separate 10 trials without the weight. And then how are we going to prove that that affects the angle dependency?}\\
\textbf{Alina}: \textit{What did you just say?}\\
\textbf{Sarah}: \textit{So, after we take the trials, how are we proving that it affects the dependency?}\\
\textbf{Alina}: \textit{Uhh, if}\\
\textbf{Sarah}: \textit{The distinguishability?}\\
\textbf{Alina}: \textit{Yeah.}\\
\textbf{Sarah}: \textit{Okay, so if t' is small, then it doesn’t affect it, but if it’s large it does.} 
\end{quote}
There is some evidence in this conversation that the group wants to prove that their mechanism is correct (e.g., ``How are we going to prove that that affects the angle dependency?"). Additionally, there is evidence they will be open to other outcomes depending on their statistical results (e.g., ``So if t' is small, then it doesn't affect it, but if it’s large it does."). These two pieces of evidence suggest the group is in the High Agency-Confirmation frame because they have a confirmatory prediction for their experiment but seem open to drawing a conclusion supported by their data.

The group collects their data and has a short interaction with the TA. They had not observed an angle dependence in the previous part of the session, so they ask the TA how they should proceed: 
\begin{quote}
\textbf{Alina}: \textit{We got a really small t', so we were like, do we need to use someone else’s data for the next part?}\\
\textbf{TA}: \textit{If you never-- so yeah you got a really small t', so it’s hard for you to be confident that you know how big that effect size is, right, you could look at the difference in the means and sort of estimate it, right? But you never really found it to be distinguishable so it’s hard for you to say, like, `Oh, the difference in the means that we found at 10$^{\circ}$ and 20$^{\circ}$ is the effect size of the angular dependence.' Yeah, you following me? So yeah, you can use their data, you could say like, `Someone else who did have a really good t' found an angular dependence of about this big, so we want to see if the effect that we’re investigating explains that, an angular dependence of that size.'}
\end{quote}
Here, the TA tries to cue the Discovery frame (light blue bar for instructor frame in Fig.~\ref{timeline}): they clarify the claim the group is investigating and do not suggest that the students should obtain a particular outcome. Instead, the TA seems to encourage the group to consider multiple possible outcomes (e.g., ``We want to see if the effect that we’re investigating explains that, an angular dependence of that size.").

This interaction does shift the group into the Discovery frame (light blue bar for student frame in Fig.~\ref{timeline}), where they appropriately state the claim they are testing and suggest they will do the experiment, find the statistical result, and interpret it accordingly (without verbalizing a particular prediction):
\begin{quote}
\textbf{Alina}: \textit{Yeah, that’s what I was about to say, I’m like, or are we just supposed to see, okay, so we are trying to see if tension makes angle matter, like if more mass means-- Should we just do the t' specifically for the two sets of data with it like this} [points to heavier pendulum bob] \textit{and see if it’s big, bigger than our t' without the extra mass?}
\end{quote}
This framing is only brief, however, as the group soon engages in another interaction with the TA in which the TA cues the Low Agency-Falsification frame (red bar for instructor frame in Fig.~\ref{timeline}):
\begin{quote}
\textbf{Alina}: \textit{Okay, so with testing tension, we were just wondering if this makes sense, the way we were thinking of it. Where it’s like, so we added the extra mass and we did it at 10$^{\circ}$ and 20$^{\circ}$.}\\
\textbf{TA}: \textit{Why 10$^{\circ}$ and 20$^{\circ}$?}\\
\textbf{Alina}: \textit{I was like maybe we can see the t' between those two angles for when it has the extra mass and then}\\
\textbf{TA}: \textit{Sure, but this is the idea that it’s the increased tension at 20$^{\circ}$ degrees that causes the angular dependence between 10$^{\circ}$ and 20$^{\circ}$ right? So adding more mass, how is that going to change the angular dependence?}\\
\textbf{Alina}: \textit{More tension.}\\
\textbf{TA}: \textit{But have you increased the difference in tension from 10$^{\circ}$ and 20$^{\circ}$?}\\
\textbf{Alina}: \textit{No.}\\
\textbf{TA}: \textit{I’m asking, I don’t know...Remember, you don’t have to prove that tension is the source, all you have to do is prove that it isn’t.}\\
\textbf{Alina}: \textit{Okay.}\\
\textbf{TA}: \textit{It’s way easier to falsify it than to try to support it, okay?}\\
\textbf{Alina}: \textit{Okay.}\\
\textbf{TA}: \textit{If you think of it in that way, you say, if the hypothesis goes that there’s a difference in period between 10$^{\circ}$ and 20$^{\circ}$ primarily because there’s a difference in tension at 10$^{\circ}$ and 20$^{\circ}$, there’s definitely ways you could go about saying `No, differences in tension don’t necessarily cause differences in period to the extent that we saw at diff-- at those various angles.' So, if you remember that, like you’re trying to falsify it, this becomes a way easier task.}
\end{quote}
Upon hearing the group's experimental design, the TA responds with similar moves to when Group A presented their confirmatory experimental design (described above). For example, the TA provides suggestions to redirect the group's experimental design (e.g., ``Why 10$^{\circ}$ and 20$^{\circ}$?") and mentions that falsifying a claim is easier than confirming a claim (e.g., ``It’s way easier to falsify it than to try to support it"). In contrast with Group A, however, the TA suggests to Group B that falsification is the correct or desired result (e.g., ``You don’t have to prove that tension is the source, all you have to do is prove that it isn’t.") and therefore cues the Low Agency-Falsification frame.  Following this interaction, Group B takes up the instructor’s cues and remains in the Low Agency-Falsification frame for the rest of the activity (red bar for student frame in Fig.~\ref{timeline}), which we detail in our prior work~\cite{sundstrom2022perc}.

In summary, the TA cues the Low Agency-Falsification frame for Group B using similar moves as with Group A, despite Group B taking up the Discovery frame before the TA interaction and not verbalizing to the TA any expectations of confirmation. Similar to Group A, these TA moves prompt Group B to shift to the Low Agency-Falsification frame.

\section{Discussion}
\label{sec:discussion}

We performed a video analysis of one nontraditional lab session to identify ways in which the students epistemically framed the activity and the ways in which the instructor cued student framing. In the following sections, we relate the set of epistemic frames we identified from the video analysis to both our theoretical framework and student frames described in the research literature. We then synthesize our findings related to student frame shifts cued by the instructor.

\subsection{Broader examples of epistemic frames in nontraditional labs}

We used our theoretical framework (see Fig.~\ref{theoreticalframework}) to characterize epistemic frames based on students' engagement with experimental predictions and epistemic agency. We identified six different epistemic frames defined along these two dimensions: whether or not students hold a prediction for the experimental result and the extent to which students enact epistemic agency (see Fig.~\ref{frames}). These frames also distinguish predictions towards confirmation (proving a claim to be true) and falsification (disproving a claim or proving a claim to be false). 

The identified frames expand upon previous work by providing more nuanced definitions of the confirmation and hoops frames previously observed in nontraditional lab instruction and mapping these frames on to our theoretical framework~\cite{hu2017qualitative,hu2018examining,smith2018surprise,smith2020pretend,stein2018confirming,smith2020expectations}. 
In particular, we more closely distinguish these two low-agency frames as students either having an experimental prediction (confirmation) or not (hoops), and also identify differentiating student behaviors (see Fig.~\ref{frames}). For example, students who articulate a prediction and exhibit less epistemic agency (the Low Agency-Falsification or Low Agency-Confirmation frames) hedge their conclusions toward their predicted outcome, while students who do not articulate a prediction and exhibit less epistemic agency (the hoops frame) draw conclusions with little to no discussion.

Our previous work on framing in nontraditional labs~\cite[e.g.,][]{phillipsnotengaging, smith2020expectations} has focused on understanding these confirmation and hoops frames because students in these frames often do not engage in the intended scientific practices~\cite{phillipsnotengaging}. Our video evidence, however, substantiates three additional frames that \emph{do} reflect student engagement in the intended scientific practices. Students in these three new frames -- High Agency-Falsification, High Agency-Confirmation, and Discovery -- enact epistemic agency to construct knowledge regardless of whether they initially articulate a predicted result. We observed students in these frames designing an experiment capable of testing the claim under investigation, reacting to data or results with questions, and iterating on the experimental design. 

We also found that students articulating a prediction for their experimental result did not necessarily take up low-agency frames (particularly confirmation) as prior literature suggests~\cite{smith2020expectations,phillipsnotengaging}. For example, we observed students verbalize a prediction and then enact epistemic agency to revise this prediction in the face of new experimental evidence (e.g., Group F). Furthermore, another study has also suggested that confirmation framing can be productive for student sensemaking as students effectively troubleshoot their experiment and models when their data disagree with their predictions~\cite{descamps2022perc}. 

Together, our evidence substantiates previous studies that argue for supporting students' epistemic agency in order for them to engage productively in scientific practices~\cite{miller2018addressing,stroupe2014examining,zwickl2015reasoning}. We also add perspectives on how students' epistemic agency relates to how they are framing the activity. We find that students' productive engagement (evidenced by the observed behaviors listed in Fig.~\ref{frames}) is independent of whether they articulate predictions for the experimental result. Thus, one might expect that an instructor should seek to shift students' epistemic agency, rather than their experimental predictions, in order to shift their overarching epistemic framing. Our analysis of the TA in this study supports this inference, as we discuss next.

\subsection{Role of the instructor in cuing student framing}

The lab instructor in our data mostly attended to students' experimental predictions, intending to shift students out of the confirmation frame by introducing the notion of falsification: that an experiment should seek to disprove a claim rather than to try to prove it to be true (see Section~\ref{sec:interview}). The instructor's cues led to five of the seven groups taking up some form of falsification framing and only one group being consistently in the Low Agency-Confirmation frame throughout the activity (Group E). The presented episodes also serve as supportive evidence that responsive teaching~\cite{robertson2016responsive} from a lab instructor has the potential to shift student frames. Responsive teaching~\cite{russ2009making,coffey2011missing,hammer2012responsive,levin2012becoming} involves understanding individual students' expectations for the activity, determining whether and how this framing is aligned with the goals of the lab, and either stabilizing or shifting this framing accordingly. With one group (Group A), for example, the instructor heard the students design a confirmation experiment, implemented a few strategies to shift them to a new frame (namely, introducing the notion of falsification), and their prediction shifted from confirmation to falsification. This ability of the instructor to intentionally shift student frames in real time has previously been demonstrated in physics problem solving contexts~\cite{modir2017students,irving2013transitions,thompson2016algorithmic,chari2019student}, but not labs~\cite{phillipsnotengaging} as we observe here. 


The instructor's cues to shift students' frames were less successful, however, when the cues were not responsive to students' current frames. Often, these cues sought to shift students out of expecting a confirmatory result whether or not they were in a frame related to confirmation. For example, the instructor interacted with Group B as if they were in a confirmation frame even though they were initially in the Discovery frame. After the instructor made a bid to shift to falsification, the group shifted towards falsification and with less epistemic agency. 

Based on the interview with the instructor, we expect that the role of experimental predictions and epistemic agency for students' frames were possibly conflated in the instructor's intentions and behaviors. 
We see this when the instructor's attempts to confront low epistemic agency (i.e., avoid students seeking to confirm existing knowledge rather than generate new knowledge) resulted in shifts in students' experimental predictions (between confirmation or falsification or no prediction), with either little or negative change to students' epistemic agency. The opposite was also true; attempts to shift students away from focusing on a particular prediction (particularly confirmation) resulted in shifts in students' epistemic agency. For example, with two of the lab groups (Groups A and B), the instructor provided a lot of guidance on their experimental designs and cued less epistemic agency (e.g., telling students what they ``have to do" to complete the assignment) in an attempt to relieve them from particular experimental predictions. These behaviors shifted students towards falsification, as intended, but also shifted students into low-agency frames. In this case, the instructor was also seeking to balance supporting students' agency over their experimental investigation (i.e., ``scientific" uncertainty) with lowering student agency over the assignment itself (i.e., ``student'' uncertainty)~\cite{manz2018supporting}. The evidence here demonstrates the tensions in structuring and supporting multiple forms of student agency, while simultaneously attending to student framing. Alternatively, the evidence may demonstrate that students had an easier time responding to instructor bids to shift their predictions than bids to shift their epistemic agency.

Although we focused our analysis on the role of the instructor, the lab instructions also likely impacted both students' epistemic framing and the instructor's behaviors (Fig.~\ref{theoreticalframework}). As we noted earlier, the activities guided by the instructions placed emphasis on examining discrepancies in data, whether or not that discrepancy was evident in the students' data. In the video data we analyzed, the only group who found the discrepancy for themselves earlier in the session was also the only group to take up a high-agency frame for the whole activity (Group G). In contrast, another group (Group B) shifted frames multiple times at the beginning of the activity: their data did not suggest an angle dependence and the group did not understand the role of the mechanism testing activity in their broader knowledge construction process. The instructor also constructed their own intent from the lab instructions (namely, introducing falsification) and similarly attempted to confront confirmation framing even if students were in a different frame. Future work should further investigate the ways in which the lab instructions impact both student framing and instructor behaviors and how static lab instructions might attend and be responsive to student framing.

Overall, these results speak to the many challenges for TAs, instructors, and curriculum developers in shifting student frames in nontraditional labs. As in many instructional labs, the TA in this video data had limited time with each individual group as they sought to address the needs of seven lab groups all conducting their own investigations. Despite this and other challenges, the instructor attended to student framing and, in many cases, the cues for frame shifts were successful (e.g., the number of groups shifting from confirmation to falsification).

\section{Conclusion}

We conducted a video analysis of an early-semester, nontraditional lab session to identify students' epistemic frames and determine the role of the instructor in shifting students between these frames. We defined and distinguished six epistemic frames in this context along two dimensions: whether or not students have a prediction for their experimental result and the extent to which students enact epistemic agency. Results suggest that attending to student agency may prompt more productive engagement in scientific practices than attending to students' experimental predictions. Our analysis of the instructor's interactions with individual lab groups also suggests that instructor cues for frame shifts can be successful when highly focused, but may be unsuccessful (or lead to unintended shifts) if they are not responsive to students' current framing. This study elucidates the complexities of student epistemic framing and the challenges of instructor attempts to shift that framing, warranting similar analyses of other nontraditional lab contexts in future research.  



\section*{ACKNOWLEDGEMENTS}

This material is based upon work supported by the NSF GRFP Grant No. DGE-2139899 and NSF Grant No. DUE-2000739. Additional support of AMP was provided by a Cottrell Postdoctoral fellowship from the Research Corporation for Science Advancement. We also acknowledge support from collaborators Ian Descamps and Sophia Jeon.

\section{APPENDIX}

\subsection{Instructor interview protocol}

\begin{enumerate}
\itemsep0em 
    \item What is your previous teaching experience, both before and at Cornell?
    \vspace{-0.2cm}
    \begin{enumerate}\itemsep0em 
        \item How many times did you teach these labs?
    \end{enumerate}
    \item In general, what was your experience teaching these labs like? 
    \vspace{-0.2cm}
    \begin{enumerate}\itemsep0em 
        \item What did you enjoy?
        \item What were some challenges?
    \end{enumerate}
    \item  What is your understanding of the learning goals of the pendulum lab?
    \vspace{-0.2cm}
    \begin{enumerate}\itemsep0em 
        \item What is your understanding of the learning goals of the mechanism testing activity? 
        \item How did you approach teaching this activity?
    \end{enumerate} 
   \item What did you mean by falsification in this context?
   \vspace{-0.2cm}
   \begin{enumerate}\itemsep0em 
        \item Why did you use this idea of falsification? 
    \end{enumerate} 
\end{enumerate}

\bibliography{framing.bib}

\end{document}